\documentclass[aps,prb,12pt,showkeys,superscriptaddress,showpacs,reprint,longbibliography]{revtex4-1} 
\usepackage{amsmath}
\usepackage{amsthm}
\usepackage{graphicx}
\usepackage[colorlinks]{hyperref}
\usepackage{dcolumn}
\allowdisplaybreaks
\usepackage[usenames,dvipsnames,table]{xcolor}
\usepackage{fancyhdr}
\usepackage{mathptmx}
\usepackage[small,raggedright]{titlesec}
\usepackage{wrapfig}
\begin{document}
\thispagestyle{plain}
\title
{
\textcolor{BlueViolet}
{
Infinite-order diagrammatic summation approach to explicitly correlated congruent 
transformed Hamiltonian
}
}
\thispagestyle{plain}
\author{Mike Bayne}
\affiliation{Department of Chemistry, Syracuse University, Syracuse, New York 13244 USA}
\author{John Drogo}
\affiliation{East Syracuse Minoa High School, Syracuse, New York 13057 USA}
\author{Arindam Chakraborty}
\email[corresponding author: ]{archakra@syr.edu}
\affiliation{Department of Chemistry, Syracuse University, Syracuse, New York 13244 USA}
\keywords{explicitly correlated, congruent transformed Hamiltonian, resolution of identity, partial infinite order summation, Gaussian-type geminal, diagrammatic summation}
\pacs{31.15.V}
\begin{center}
\begin{abstract}
\noindent\textcolor{BlueViolet}{\rule{14cm}{1.2pt}}
        \colorbox{Apricot}{\parbox{13.7cm}{
\begin{wrapfigure}{r}{1.35in}
\vspace{-35pt}
\includegraphics[width=1.9in]{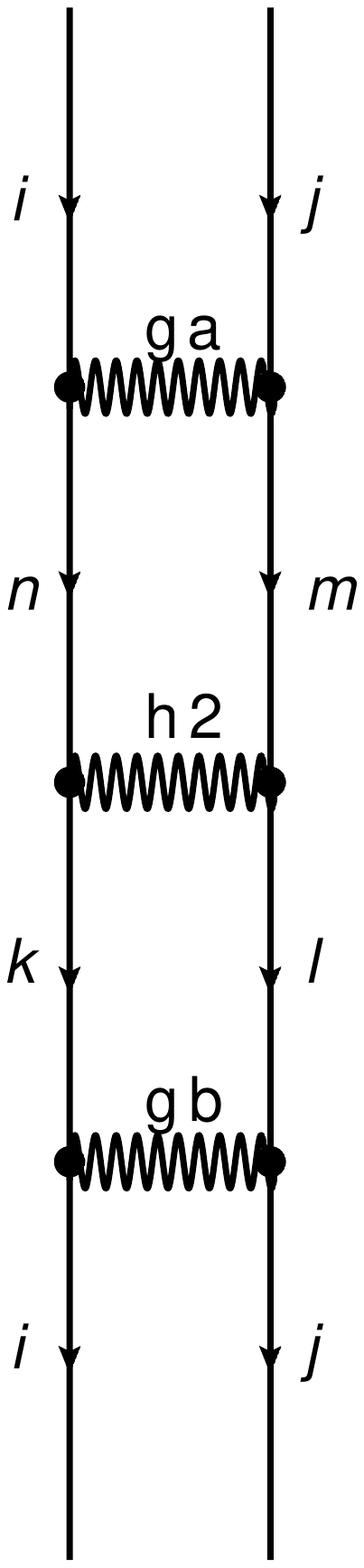}
\vspace{-70pt}
\end{wrapfigure}
A resolution of identity approach to explicitly correlated 
congruent transformed Hamiltonian (CTH) is presented. One of the 
principle challenges associated with the congruent transformation 
of the many-electron Hamiltonian is the generation of 
three, four, five, and six particle operators. Successful
application of the congruent transformation requires efficient
implementation of the many-particle operators. In this work,
we present the resolution of identity congruent transformed Hamiltonian (RI-CTH)
method to handle many-particle operators. 
The resolution of identity was used to project the explicitly correlated operator
in a $N$-particle finite basis to avoid explicit computation of the many-particle operators. 
Single-particle states were obtained by performing Hartee-Fock calculations, which 
were then used for construction of many-particle states. The limitation of the 
finite nature of the resolution of identity was addressed by
developing partial infinite order (PIOS) diagrammatic summation 
technique. In the PIOS method, the matrix elements of the projected congruent transformed Hamiltonian was expressed
in terms of diagrammatic notation and a subset of diagrams were summed up to infinite order.
The RI-CTH and RI-CTH-PIOS methods were applied to isoelectronic series of 10-electron systems
$(\mathrm{Ne,HF,H_2O,NH_3,CH_4})$ and results were compared with CISD and CCSD(T) calculations.
One of the key results from this work is that for identical basis set, the RI-CTH-PIOS energies 
are lower than CISD and CCSD(T) values. 
}
}
\noindent\textcolor{BlueViolet}{\rule{14cm}{1.2pt}}
\end{abstract}
\end{center}
\maketitle
\textcolor{BlueViolet}{\section{Introduction} \label{sec:intro}}
The form of the many-electron wavefunction at small electron-electron separation 
plays an important role 
in accurate determination of the ground state energy. The relationship between  the 
Coulomb singularity in the electronic Hamiltonian and form of the many-electron wavefunction 
at the electron-electron coalescence point is
well known and is given by the 
Kato cusp condition.~\cite{kato1957eigenfunctions,hammond1994monte,lester1997recent,nightingale1998quantum}
Explicitly correlated methods 
improve the form of the many-electron wavefunction near the 
electron-electron coalescence point by 
incorporating explicit $r_{12}$ dependence in the form of 
the wavefunction. This approach has been shown to 
be indispensable  for high-precision calculations of ground and excited 
state energies in atoms and molecules and 
has been implemented in various methods including 
quantum Monte Carlo(QMC),~\cite{hammond1994monte,lester1997recent,nightingale1998quantum,clark2011computing,wagner2003quantum} 
perturbation theory (MP2-R12),~\cite{manby1999real,manby2006explicitly,valeev2006combining,
shiozaki2010communications} coupled-cluster,~\cite{lane2012ccsdtq,
shiozaki2009explicitly,valeev2008simple,kohn2010explicitly,
hattig2010communications,kohn2009explicitly,shiozaki2009higher,parkhill2011formulation}
configuration interaction,
transcorrelated Hamiltonian,~\cite{boys1969condition,boys1969determination,yanai2012canonical,
ten2000feasible,umezawa2003transcorrelated,hino2002application,umezawa2004ground} and geminal 
augmented MCSCF.~\cite{varganov2010variational} One of the 
main challenges in efficient implementation of explicitly correlated methods is the 
analytical evaluation of integrals involving the $r_{12}$ term. The electronic 
Hamiltonian has only one and two-particle operators, however, because of the 
$r_{12}$ term in the form the wavefunction, integrals involving the Hamiltonian 
and explicitly correlated wavefunctions often involve three-particle and higher terms.
The resolution of identity (RI) approach 
has been successful for efficient evaluation of many-particle integrals and
has been widely adopted for
implementing faster, more efficient R12-MP2~\cite{klopper2002explicitly,tew2007weak,
ten2003density,valeev2004improving,lambrecht2011kinetic} and R12-CC methods.~\cite{shiozaki2008explicitly}
\par
In this article, we introduce the RI implementation of the 
explicitly correlated congruent transformed Hamiltonian (CTH) method.~\cite{elward2012variational} 
In the CTH method an explicitly correlated function is used, to 
perform congruent transformation~\cite{james1968mathematics,datta2004matrix}
of the electronic Hamiltonian.
This approach is similar to the transcorrelated Hamiltonian method where a
similarity transformation is performed on the Hamiltonian.~\cite{boys1969condition,boys1969determination} 
One of the advantages of the CTH method is that the transformation preserves the 
Hermitian property of the Hamiltonian. As a consequence, the transformed
Hamiltonian is amenable to standard variational procedures for 
obtaining the ground state energy.~\cite{elward2012variational} 
The transformed Hamiltonian involves upto six-particle 
operators and efficient implementation of these many-particle operators
is crucial for application of the CTH method. This problem is addressed in the 
present work by introducing the RI approximation for representing the 
many-particle operators. The RI method is exact in the limit of the 
infinite number of basis functions, however, practical implementation of the 
RI is always approximate because of the truncation of the basis. 
Here, we present a diagrammatic 
summation approach to include infinite-order contributions to the 
finite basis implementation of the RI method. We have used 
diagrammatic notation that is commonly used in 
the perturbation theory and coupled-cluster
equations to represent the terms in the RI expansion.~\cite{shavitt2009many} After that, we show 
that certain classes of diagrams can be summed upto infinite-order and the 
result can be expressed as an analytical expression of a renormalized 
two-particle operator. Because the method in its current form is applicable only to selected 
(as opposed to all) classes of diagrams, it is denoted as partial infinite-order 
summation (PIOS) method. The details of the derivation of the PIOS method are 
presented in the following section. The PIOS method has been used for calculating 
the ground state energy of isoelectronic 10-electron systems
$(\mathrm{Ne,HF,H_2O,NH_3,CH_4})$ and results are presented in \autoref{sec:results}.

\textcolor{BlueViolet}{\section{Theory} \label{sec:theory}}
\subsection{Resolution of identity}
The first step in the construction of the CTH is to define an explicitly correlated 
two-body operator as shown below
\begin{align}
	G(1,\dots,N) = \sum_{i<j}^{N} g(r_{ij}) =  \sum_{i<j}^{N} g(i,j),
\end{align} 
where $N$ is the number of electron in the system. The derivation presented here 
is independent of the choice of the two-body explicitly correlation function 
$g(1,2)$ and the specific form used in the present calculation will be discussed later.
The congruent-transformed operator is defined as
\begin{align}
	\tilde{H} &= G^\dagger H G \\
	\tilde{S} &= G^\dagger \mathbf{1} G
\end{align}
where the transformed Hamiltonian contains upto six-particle operators.~\cite{elward2012investigation,elward2012variational}
For a given trial wavefunction $\Psi_\mathrm{T}$, the CTH energy is defined as
\begin{align}
\label{eq:ecth}
	E[\Psi_\mathrm{T},G] = \frac{\langle \Psi_\mathrm{T} \vert \tilde{H} \vert \Psi_\mathrm{T} \rangle}{\langle \Psi_\mathrm{T} \vert \tilde{S} \vert \Psi_\mathrm{T} \rangle}.
\end{align}
The congruent transformation preserves the Hermitian property of the electronic Hamiltonian 
and by construction the CTH energy is an upper bound to the exact ground state energy
\begin{align}
	E_\mathrm{exact} 
	\le \min_{\Psi_\mathrm{T},G} E[\Psi_\mathrm{T},G] 
	\le \min_{\Psi_\mathrm{T}} E[\Psi_\mathrm{T},G=1].
\end{align}
As a consequence of the above relationship, the CTH energy is amenable to standard variational procedure and can 
be minimized with respect to both the trial wavefunction $\Psi_\mathrm{T}$ and the 
explicitly-correlated function $G$. In the limit of $G=1$, the CTH energy is equivalent 
to the expectation value of the electronic Hamiltonian. One of the challenges 
of implementing CTH is form of the transformed Hamiltonian. Because of the transformation, the
CTH can be expressed as sum of two, three, four, five, and six-particle operators. In this work,
we address this challenge by introducing a finite-basis for representing the CTH.  
The resolution of identity operator, in some finite basis $M$ is define by the following equation
\begin{align}
\label{eq:resofid}
  {I}^{\mathrm{RI}(M)} = \sum_{k}^{M} \vert k\rangle\langle k \vert.
\end{align}
The finite-basis representation of $G$ is given as 
\begin{align}
\label{eq:gm}
 G^{(M)}(1,\dots,N) = \sum_{k,k'}^{M}
                  \vert \Phi_k \rangle \langle \Phi_k \vert 
                   G 
                 \vert \Phi_{k'} \rangle \langle \Phi_{k'} \vert,
\end{align}
where the superscript $M$ in $G^{(M)}$, represents that it is a
finite-basis representation of the exact $G$ operator. These two
quantities are related to each other by the following limiting condition
\begin{align}
	G = \lim_{M \rightarrow \infty} G^{(M)}.
\end{align}
The number of terms in Eq. \eqref{eq:gm} that contribute to the CTH energy in Eq. \eqref{eq:ecth} is much less than $M^2$  and 
depend on the choice of the trial wavefunction $\Psi_\mathrm{T}$. If the 
search for the optimal trial wavefunction is restricted to the set of single Slater 
determinants, then as a direct consequence of Slater-Condon rules,~\cite{szabo1989modern}
the terms in the expansion are restricted to only singles and doubles excitation.
In the following equation, the notation 
$k \in S,D$ and $G^\mathrm{(\mathrm{S,D})}$ is used to denote that the only singles and 
doubles are included in the expansion 
\begin{align}
\label{eq:em}
	E[\Phi_0,G^\mathrm{(\mathrm{S,D})}] = \frac{\sum_{k,k' \in S,D} G_{0k} H_{kk'} G_{k'0} }
	{\sum_{k \in S,D} G_{0k} G_{k0} },
\end{align}
where $G_{kk'}$  and $H_{kk'}$ are shorthand notation for matrix elements 
$\langle \Phi_{k}\vert G \vert \Phi_{k'} \rangle$ 
and $\langle \Phi_{k}\vert H \vert \Phi_{k'} \rangle$, respectively. 
The ground state energy is obtained variationally by 
minimizing the total energy with respect to the geminal parameters and the Slater determinant
as shown below
\begin{align}
 E_\mathrm{RI-CTH} = \min_{g,\Phi_0} E[\Phi_0,G^\mathrm{(\mathrm{S,D})}].
\end{align}
\subsection{Form of the correlation function}
Although the expression in Eq. \eqref{eq:em}
is valid for any form of $g(1,2)$, the computational cost and ease of implementation 
depends on the specific choice of $g(1,2)$. In this work, we have used Gaussian-type
geminal (GTG) functions~\cite{persson1997molecular,bukowski1999gaussian,ten2000feasible,ten2003explicitly,manby2006explicitly,valeev2006combining,Tenno2007175,dahle2007accurate,valeev2008simple,varganov2010variational,yanai2012canonical} for representing the 2-body correlation function
\begin{align}
	g(r_{12}) = \sum_{k=1}^{N_\mathrm{g}} b_k e^{-r_{12}^2/d^2_k} ,
\end{align} 
where $b_k,d_k$ are the geminal parameters that completely define the GTG function. 
There are mainly two different techniques for determining the geminal parameters. 
In the first method, the parameters are determined variationally by minimizing the total energy. Although this approach 
very accurate, it becomes computationally expensive because it involves
multidimensional minimization and recomputation of the atomic orbital (AO) integrals.   
The second approach is to have a set of precomputed values of the geminal parameters. 
This approach is computationally fast, however, the challenge it to find a transferable
set of parameters that can be applied to various molecules.
In this work, we have
developed a mixed approach where
the linear geminal parameters $b_k$ are variationally
optimized during the calculation and the non-linear geminal parameters $d_k$ are precomputed before the start of the geminal optimization.
 The central ideal of this method 
is to use some appropriate characteristic length scale associated with the 
molecule for calculating the $d_k$  parameters. We have used the average electron-electron separation distance 
as the characteristic system-dependent quantity for calculating 
the geminal parameters. Using the reference Slater determinant $\Phi_0$, 
we define the average electron-electron separation as  
\begin{align}
	\langle r^2_{12} \rangle_0 
	&= \frac{1}{N(N-1)} \langle \Phi_0 \vert \sum_{i<j} r_{ij}^2 \vert \Phi_0 \rangle .
\end{align}
The $d_k$ parameters are selected from a set of numbers obtained by scaling
$\langle r^2_{12} \rangle_0$
\begin{align}
\label{eq:dk}
 d_k^2 \in 
 \left[
 \frac{1}{n}\langle r^2_{12} \rangle_0,
 \dots
 \frac{1}{2}\langle r^2_{12} \rangle_0,
 \langle r^2_{12} \rangle_0, 2\langle r^2_{12} \rangle_0,
 \dots n\langle r^2_{12} \rangle_0 
 \right].
\end{align}
The choice of $\langle r^2_{12} \rangle_0$ over 
$\langle r_{12} \rangle_0$ was made purely for computational convenience.
The integral involving $r_{12}^2$ is separable in $x,y,$ and $z$
components and can be integrated easily with Cartesian Gaussian-type 
orbitals (GTOs). Similar separation is not possible for $\langle r_{12} \rangle_0$.
The above procedure provides a fast and physically intuitive method for
obtaining the non-linear geminal parameters. 
\par
After the non-linear $d_k$ parameters were obtained using the steps described above,
the linear geminal parameters $b_k$ were optimized variationally. We have 
avoided recomputation of the AO integrals by postponing the inclusion of
the $b_k$ terms to the very last step of the calculation. This is shown 
by the following example. The expectation value of the geminal operator
is defined as
\begin{align}
\label{eq:pmunu}
	\langle \Phi_0 \vert G \vert \Phi_0 \rangle
	=
	\sum_{\mu \nu \lambda \sigma}
	P_{\mu \nu} P_{\lambda \sigma}
	\left(
	\frac{1}{2}[\mu \nu \vert g \vert \lambda \sigma]
	-\frac{1}{4} [\mu \sigma \vert g \vert \lambda \nu] 
	\right) 
\end{align}
where $\mu,\nu,\lambda,\sigma$ are AO indices, $\mathbf{P}$ 
is the density matrix, and the integrals are in chemist's notation.~\cite{szabo1989modern} 
Substituting the expression for the geminal function,
Eq. \eqref{eq:pmunu} can be written as
\begin{align}
	\langle \Phi_0 \vert G \vert \Phi_0 \rangle
	&= 
	\sum_{k}^{N_\mathrm{g}}
	b_k
	\sum_{\mu \nu \lambda \sigma}
	P_{\mu \nu} P_{\lambda \sigma} A^k_{\mu \nu \lambda \sigma} \\
\label{eq:ak}
	A^k_{\mu \nu \lambda \sigma}
	&=
	\frac{1}{2}[\mu \nu \vert e^{-r_{12}^2/d_k^2} \vert \lambda \sigma]
	-\frac{1}{4} [\mu \sigma \vert e^{-r_{12}^2/d_k^2} \vert \lambda \nu] .
\end{align}
The quantity $A^k_{\mu \nu \lambda \sigma}$ is independent of
$b_k$ and was computed once and stored at the start of the $b_k$
optimization.  
\par
One of the advantages of the GTG
function is that the AO integrals involving the GTG functions are
analytical and can be expressed in a closed form.
Analytical expressions for integrals involving 
s-type GTO are known and 
were derived by Boys.~\cite{boys1960integral} Analytical form for the higher
angular momentum GTOs using Mcmurchie-Davidson algorithm was derived by Persson and Taylor.~\cite{persson1997molecular}
Because of the availability of fast analytical integral routines,  Gaussian-type geminal functions 
have found widespread application in a large number of explicitly correlated calculations.~\cite{persson1997molecular,bukowski1999gaussian,ten2000feasible,
ten2003explicitly,manby2006explicitly,valeev2006combining,
Tenno2007175,dahle2007accurate,valeev2008simple,
varganov2010variational,yanai2012canonical,
elward2012investigation,elward2012variational}
As seen in Eq.\eqref{eq:em}, the
geminal integrals needed for computation of the energy expression
is for the form $G_{0k}$. These geminal integrals are known as 
the overlap integrals and are especially efficient to compute because
they can be written as a product of three 1D integrals
\begin{align}
\label{eq:xyz_int}
	[\mu \nu \vert e^{-r_{12}^2/d_k^2} \vert \lambda \sigma]
	&= I_x I_y I_z .
\end{align} 
The exact expression for the integrals can be found in Refs.~\citenum{boys1960integral,persson1997molecular}
In addition to restricting the terms in Eq. \eqref{eq:em} to only singles and doubles,
the fast evaluation 
of $G_{0k}$ was used to further restrict the number of terms in the summation. 
We have implemented a ``direct'' approach in which the full $\mathbf{H}$ matrix
is never constructed and the matrix element $H_{kk'}$ 
are computed as needed during the course of the calculations. 
The evaluation of $H_{kk'}$ is only performed 
when $G_{0k}G_{0k'}$ is higher than some threshold value
\begin{align}
\label{eq:gem_limit}
  \vert G_{0k} G_{0k'} \vert > \Delta_\mathrm{tol}.
\end{align}
Overall, equations \eqref{eq:dk}, \eqref{eq:ak}, \eqref{eq:xyz_int}, and \eqref{eq:gem_limit}
represent the four key steps for efficient implementation of the RI-CTH methods.  
\subsection{Partial infinite-order summation}
\label{sec:pios}
\begin{figure}[h!]
  \begin{center}
    \includegraphics[width=80mm,height=60mm]{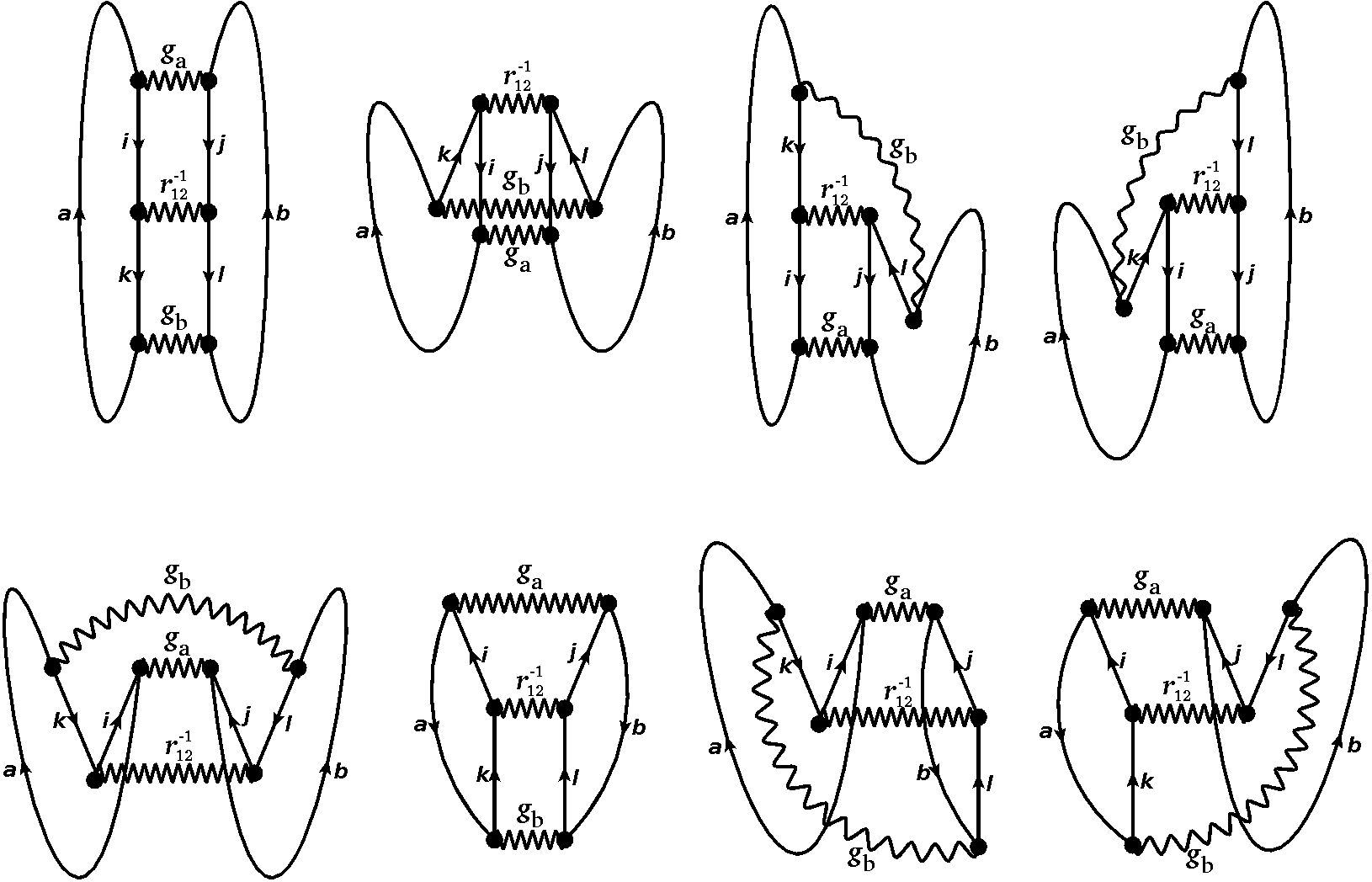}
    \includegraphics[width=80mm,height=60mm]{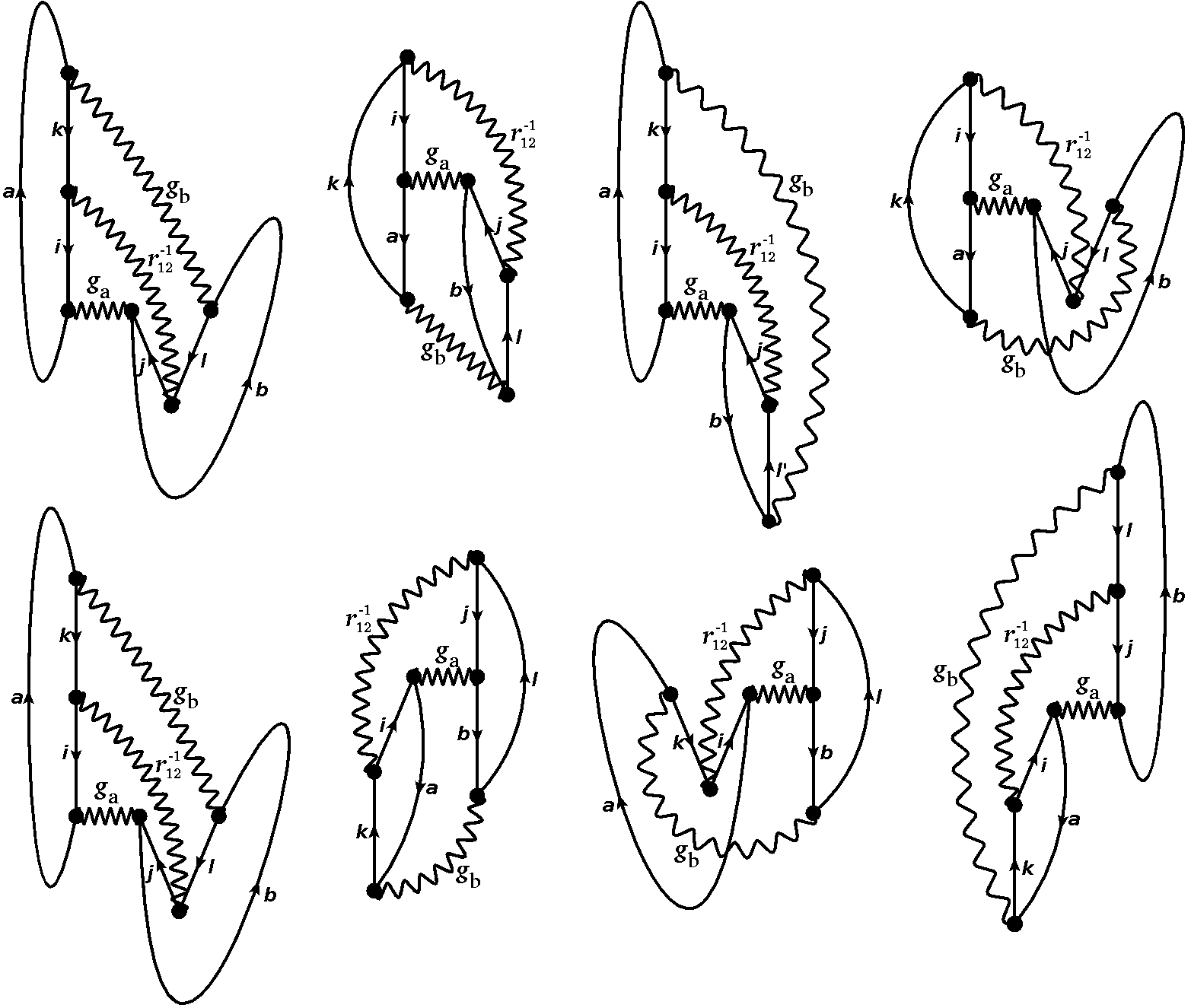}
    \caption{Coulomb diagrams that are summed to infinite order}
    \label{fig:GS1}
  \end{center}
\end{figure}

Up to this point, only finite expansion of the RI-CTH method has been considered. In this 
section, we will develop the infinite order summation approach. The main idea
of this approach can be summarized in two steps. In the first step, the 
RI-CTH energy terms were expressed in terms of diagrammatic notations. 
In the next step, we used the diagrammatic summation technique to perform 
infinite order summation for certain classes of diagrams.  
Starting with Eq.~\eqref{eq:em}, we define the 2-particle transition density matrix as
\begin{align}
	\Gamma_{ijkl}^{uv} = \langle \Phi_u \vert i^\dagger j^\dagger l k \vert \Phi_v \rangle .
\end{align}
Using the above expression, the numerator in the Eq.~\eqref{eq:em}
can be written as 
\begin{align}
	G_{0m} 
	= \sum_{ab}^{N_\mathrm{occ}} \sum_{ij} 
	\langle ab \vert g \vert ij \rangle  \Gamma_{abij}^{0m}
\end{align}
where, the indices $i,j,k,l...$ are used for labeling the molecular orbitals (MOs).
We have used the convention~\cite{szabo1989modern}  of labeling the MOs that are occupied and unoccupied
in the reference Slater determinant as $a,b,c,d...$ and $p,q,r,s...$,
respectively.  
The overall expression of the electron-electron interaction in the RI-CTH energy is given as 
\begin{align}
\label{eq:vee}
	G_{0m} V_{mm'}^\mathrm{ee} G_{m'0}
	&= \sum_{abcd}^{N_\mathrm{occ}} \sum_{ijkl}
    \Gamma_{abij}^{0m} \Gamma_{i'j'k'l'}^{mm'}  \Gamma_{klcd}^{0m'} \\ \notag
	&\times \langle ab \vert g \vert ij \rangle  
	\langle i'j' \vert r_{12}^{-1} \vert k'l' \rangle  
	\langle kl \vert g \vert cd \rangle.
\end{align}
In the diagrammatic representation, the 
occupied MO indices $a,b,c,d$ are represented 
by hole lines $(\downarrow)$. The general MO indices ${ijkli'j'k'l'}$
can be either particle $(\uparrow)$ or hole $(\downarrow)$ lines .
For a finite RI expansion, the energy expression can be expressed in terms 
of finite number of diagrams. After careful analysis of the diagrams, we 
selected a subset of diagrams that were summed up to infinite order. 
The objective of performing 
the diagram summation is to obtain a compact renormalized 
operator that is more computationally tractable than the explicit 
infinite-order sum. The selection of diagrams 
for summation was based on the ease of implementation of the 
resulting renormalized operator. Since only a subset of diagrams (as oppose
 to all) were selected for summation
upto infinite order, we denote this method as partial infinite-order summation (PIOS) method.
The Coulomb diagrams that were summed upto infinite order in the PIOS method
are presented in \autoref{fig:GS1}.   
The summation of diagrams leads to the following expression 
\begin{align}
\label{eq:renorm}
  \sum'{\mathrm{diagrams}} = \frac{\langle\Phi_{\mathrm{HF}}|g(1,2)r_{12}^{-1}g(1,2)|\Phi_{\mathrm{HF}}\rangle}
                                 {\langle\Phi_{\mathrm{HF}}|g(1,2)g(1,2)|\Phi_{\mathrm{HF}}\rangle}.
\end{align}
which is a renormalized 
2-body operator. The prime over the summation in  Eq.~\eqref{eq:renorm}
is used to denote that only selected diagrams were included in the summation.  
Because the energy expression in Eq. \eqref{eq:em} includes a denominator, 
identical procedure was also used for obtaining the denominator and the combined 
result is shown in Eq.~\eqref{eq:renorm}. We define the 
RI-CTH-PIOS energy as
\begin{align}
  E_{\mathrm{RI-CTH-PIOS}} = \tilde{E}_{\mathrm{RI-CTH}} + E_{\mathrm{PIOS}},
\end{align}
where $E_{\mathrm{PIOS}}$ is given by Eq. \eqref{eq:renorm}. The tilde on $\tilde{E}_{\mathrm{RI-CTH}}$ is used to denote that RI-CTH energy should exclude diagrams that have been included in the PIOS energy calculation to prevent double counting.
\subsection{Computational details}
All the calculations were performed using $N_\mathrm{g}=2$ with two Gaussian-type geminal functions . The 
first set of geminal parameter were fixed at $b_1=1$ and $d_1^2=\infty$. 
This choice of parameters ensured that the RI-CTH energy is always bounded from top by the Hartree-Fock energy. 
Hartree-Fock calculation was performed and  $\langle r^2_{12} \rangle_0$ was 
evaluated. The $\langle r^2_{12} \rangle_0$ was used to construct the
following  trial set for the selection of the $d_2$ parameter
\begin{align}
 d^2_\mathrm{trial} =
 \frac{1}{3}\langle r^2_{12} \rangle_0,
 \frac{1}{2}\langle r^2_{12} \rangle_0,
 \langle r^2_{12} \rangle_0, 2\langle r^2_{12} \rangle_0,
 3\langle r^2_{12} \rangle_0 .
\end{align}
The $b_2$ parameters was optimized for each trial $d_2^2$ and the $b_{2,\mathrm{opt}},d^2_{2,\mathrm{opt}}$
were obtained by finding the lowest energy in the trial set.

\textcolor{BlueViolet}{\section{Results and conclusion}} \label{sec:results}
The RI-CTH method was applied for computing the ground state 
energy of isoelectronic 10-electron systems, $\mathrm{Ne}$, $\mathrm{HF}$, $\mathrm{H_{2}O}$, 
$\mathrm{NH_{3}}$, and $\mathrm{CH_{4}}$, and the results are presented in \autoref{tab:ten_elec}.
\begin{table*}[ht]
  \begin{center}
   \caption{\textbf{RI-CTH-PIOS energies for isoelectronic 10-electron systems. All the values are reported in atomic units. }}
   \label{tab:ten_elec}
    \begin{tabular}{ l c c c c c c }  
     \hline
        $\textrm{Method}$  & $\textrm{Ne}$  & $\textrm{HF}$  & $\textrm{H$_2$O}$  & $\textrm{NH$_3$}$  & $\textrm{CH$_4$}$ & $\textrm{References}$  \\ \hline 
        $\textrm{HF}$             & -128.474407 & -100.002394 & -76.009999 & -56.183815 & -40.194821 \\
        $\textrm{RI\text{-}CTH/}\mathrm{6\text{-}31G^{\star}}$           & -128.605009 & -100.153397 & -76.165385 & -56.327220 & -40.313661 & \textrm{This work} \\
        $\textrm{RI\text{-}CTH\text{-}PIOS/}\mathrm{6\text{-}31G^{\star}}$ & -128.635313 & -100.271451 & -76.311230 & -56.441383 & -40.387422 & \textrm{This work} \\       
        $\textrm{CISD/}\mathrm{6\text{-}31G^{\star}}$     & -128.623340 & -100.180709 & -76.198206 & -56.361897 & -40.346897 & \citenum{johnson2010computational} \\  
        $\textrm{CCSD(T)/}\mathrm{6\text{-}31G^{\star}}$  & -128.626734 & -100.186601 & -76.205841 & -56.369520 & -40.353006 & \citenum{johnson2010computational} \\  
        $\textrm{CISD/}\mathrm{cc\text{-}pV(T+d)Z}$       & -128.791918 & -100.322996 & -76.313875 & -56.454729 & -40.422120 & \citenum{johnson2010computational} \\
        $\textrm{CCSD(T)/}\mathrm{cc\text{-}pV(T+d)Z}$    & -128.798209 & -100.331994 & -76.324556 & -56.465536 & -40.431821 & \citenum{johnson2010computational} \\
      \hline
    \end{tabular}
  \end{center}  
\end{table*}   
As expected, the RI-CTH energy is much lower than the HF energies. 
It was found that both CISD and CCSD(T) energies obtained using identical basis set are lower than 
the RI-CTH energy. This is an expected result because in the CISD calculation
unconstrained optimization of the CI coefficients is performed. On the other hand,
the coefficients in RI-CTH method are constrained by the functional form of the 
geminal function. Comparing the energies from the RI-CTH-PIOS calculations with 
CISD/6-31G* and  CCSD(T)/6-31G* show that the RI-CTH-PIOS energy is consistently
lower for all the 10-electron systems. We attribute this 
lower energy to the inclusion of diagrams that were missing in the RI-CTH method but were included
because of the diagrammatic summation in RI-CTH-PIOS. The terms that are missing 
from the CISD/6-31G* calculation can be systematically included by increasing the 
size of the underlying 1-particle basis. To investigate this further, we have 
compared the  RI-CTH-PIOS energies with CISD calculation with a much larger basis set. As shown in  \autoref{tab:ten_elec}, the CISD/cc-pV(T+d)Z 
are consistently lower than the RI-CTH-PIOS energy. The results in \autoref{tab:ten_elec} indicate that the RI-CTH-PIOS method
with a small basis set is able to capture 
part of the electron correlated energy that is only accessible to CISD and CCSD(T)
methods at larger basis sets. 
\par
In conclusion, the resolution of identity implementation of 
congruent transformed Hamiltonian has been presented. The congruent 
transformation of the many-electron Hamiltonian was performed
using Gaussian-type geminal functions. The challenge of 
efficient optimization of the geminal function was addressed by 
using different strategies for optimizing linear and non-linear
parameters. The linear geminal parameters were obtained variationally 
by minimizing the RI-CTH energy. The expectation value of the square of the electron-electron separation 
distance was used as the 
characteristic length scale for construction of the non-linear geminal parameters.
One of the key results in this work is the development and application of partial infinite order summation method. Diagrammatic notation of the RI-CTH expression was introduced and the RI-CTH-PIOS calculations were performed. It was found that for identical basis functions, the  RI-CTH-PIOS energies are lower than the CISD and CCSD(T) energies for the isoelectronic 10-electron system studied in this work.

\begin{acknowledgments}
We gratefully acknowledge the support from Syracuse
University for this work.
\end{acknowledgments}
\bibliography{ricthpios}
\end{document}